\documentclass[aps, reprint, amsmath, amssymb, twocolumn]{revtex4-1}

\usepackage{xr}
\makeatletter

\newcommand*{\addFileDependency}[1]{
\typeout{(#1)}
\@addtofilelist{#1}
\IfFileExists{#1}{}{\typeout{No file #1.}}
}\makeatother

\newcommand*{\myexternaldocument}[1]{%
\externaldocument{#1}%
\addFileDependency{#1.tex}%
\addFileDependency{#1.aux}%
}

\myexternaldocument{supporting}

\makeatletter
\def\@email#1#2{%
 \endgroup
 \patchcmd{\titleblock@produce}
  {\frontmatter@RRAPformat}
  {\frontmatter@RRAPformat{\produce@RRAP{*#1\href{mailto:#2}{#2}}}\frontmatter@RRAPformat}
  {}{}
}%
\makeatother

\usepackage{graphics, graphicx, placeins}
\graphicspath{ {./media/} }
\usepackage{dcolumn}  
\usepackage[T1]{fontenc}
\usepackage{sidecap}
\usepackage{color} 
\usepackage{bm} 
\usepackage[dvipsnames]{xcolor,colortbl}
\usepackage{amsmath, amssymb, amsfonts, etoolbox}
\usepackage{hyperref}
\usepackage{cleveref}

\usepackage{xspace}
\newcommand{\ie}{\textit{i.e.}\xspace}
\newcommand{\eg}{\textit{e.g.}\xspace}

\newcommand{\insilico}{\textit{in silico}\xspace}

\newcommand{\typeA}{type A\xspace}
\newcommand{\typeB}{type B\xspace}
\newcommand{\typeAB}{type A and B\xspace}

\usepackage{epstopdf}
\AppendGraphicsExtensions{.tiff}

\usepackage[normalem]{ulem}       
\usepackage{blindtext}

\newcommand{\params}[1]{\textcolor{black}{#1}}  

\begin{document} 

\newcommand{\mytitle}{Neuroevolution of Decentralized Decision-Making in N-Bead Swimmers Leads to Scalable and Robust Collective Locomotion}


\author{Benedikt Hartl}
\affiliation{Institute for Theoretical Physics, TU Wien, Austria}
\affiliation{Allen Discovery Center at Tufts University, Medford, MA, 02155, USA}
\email{hartl.bene.research@gmail.com}
    
\author{Michael Levin}
\affiliation{Allen Discovery Center at Tufts University, Medford, MA, 02155, USA}
\affiliation{Wyss Institute for Biologically Inspired Engineering at Harvard University, Boston, MA, 02115, USA}

\author{Andreas Zöttl}
\affiliation{Faculty of Physics, University of Vienna, Austria}

\date{\today}

\title{\mytitle}
\begin{abstract}
Many microorganisms swim by performing larger non-reciprocal shape deformations that are initiated locally by molecular motors. However, it remains unclear how decentralized shape control determines the movement of the entire organism. Here, we investigate how efficient locomotion emerges from coordinated yet simple and decentralized decision-making of the body parts using neuroevolution techniques.
Our approach allows us to investigate optimal locomotion policies for increasingly large microswimmer bodies, with emerging long-wavelength body shape deformations corresponding to surprisingly efficient swimming gaits. The obtained decentralized policies are robust and tolerant concerning morphological changes or defects and can be applied to artificial microswimmers for cargo transport or drug delivery applications without further optimization ``out of the box''. Our work is of relevance to understanding and developing robust navigation strategies of biological and artificial microswimmers and, in a broader context, for understanding emergent levels of individuality and the role of collective intelligence in \textit{Artificial Life}.
\end{abstract}

\maketitle


\section{Introduction}
\label{sec:introduction}
Microorganisms are ubiquitous in nature and play an important role in many biological phenomena, ranging from pathogenic bacteria affecting our health to phytoplankton as a key player in the marine ecosystem on a global scale. 
A large variety of microorganisms live in viscous environments and their motion is governed by the physics of low Reynolds number hydrodynamics where viscous forces dominate over inertia \cite{Lauga2009a,Elgeti2015,Bechinger2016,Zottl2016}.
As a consequence, a common strategy is to periodically deform their body shape in a non-reciprocal fashion to swim.
To thrive in the environment, they have developed different tailored strategies to exploit their swimming capabilities, such as actively navigating toward a nutrient-rich source, hunting down prey, escaping predators, or reproducing \cite{Bray2000,Wan2023}.
Besides being of direct biological relevance, understanding the corresponding navigation strategies of microorganisms bears potential for biomedical or technical applications, potentially utilized by synthetic microswimmers deployed as targeted drug delivery systems~\cite{Jang2019, Singh2019, Patra2013, Kievit2011}.

Nature has evolved many different strategies and control mechanisms for microswimmers to swim fast, efficiently, and adaptive to environmental conditions \cite{Bray2000}:
For example, swimming algae cells or sperm cells move with the help of waving cilia or flagella \cite{Brennen1977}, respectively, or amoebae such as \textit{Dictyostelium} \cite{Barry2010a}  and unicellular protists such as \textit{Euglenia} \cite{Noselli2019}  by deforming their entire cell body.
Notably, the associated deformation amplitudes and wavelengths can be on the order of the entire size of the organism.

In the last years, Reinforcement Learning \cite{Sutton1998} (RL) has been applied to understand navigation strategies of microswimmers \cite{Nasiri2023,Zottl2023}, for example under external fluid flow \cite{Colabrese2017} or external fields \cite{Schneider2019}, or to perform chemotaxis in the presence of chemical gradients \cite{Hartl2021,Paz2023,Rode2024,Alonso2024,Nasiri2024}. 
So far, most of these studies treat microswimmers as rigid agents, \ie, explicitly omitting body deformations or other internal degrees of freedom, and simply manipulate their \textit{ad hoc} swimming speed and rotation rates for control purposes to perform well in a particular environment.
Only very recent contributions \cite{Tsang2020,Hartl2021,Zou2022,Qin2023d,Zou2024,Jebellat2024,Lin2024} consider the constraints of physically force-free shape-deforming locomotion of plastic model microswimmers in an effort to identify swimming gates that explicitly utilize the hydrodynamic interactions between different body-parts in a viscous environment.
Modeling such a \textit{body-brain-environment} offers explicit and more realistic behavior of the response of an organism to environmental conditions \cite{Chiel1997,Cohen2014a}.

A prominent model that has frequently been investigated with RL techniques is the (generalized) Najafi-Golestanian (NG) microswimmer \cite{Najafi2004,Earl2007,Golestanian2008}, typically consisting of $N=3$ (or more) concentrically aligned beads immersed into a viscous environment.
Such a composite $N$-bead NG microswimmer can self-propel via nonreciprocal periodic shape deformations that are induced by coordinated time-dependent periodic forces applied to every bead which sum up to zero for force-free microswimmers. 
Typical strategies to describe autonomous microswimmer locomotion utilize a centralized controller that integrates all the information about the current state of the microswimmer in its environment (comprising internal degrees of freedom, and potential environmental cues such as chemical field concentrations). As such, it proposes control instructions for every actuator in the system, thereby inducing dynamics, \ie, body deformations, that are most optimal for a specific situation given a particular task \cite{Tsang2020,Hartl2021,Jebellat2024}.
To substitute and mimic the complex and adaptable decision-making machinery of biological microswimmers, such controllers are often realized by trainable Artificial Neural Networks (ANNs).

Centralized decision-making relies on the (sensory) input of all individual body parts of a composite microswimmer, \ie, quantities such as the relative positions, velocities, or other degrees of freedom for all $N$ beads, and the corresponding control actions target all system actuators, \ie, what forces to apply to every one of the $N$ beads.
While the number of trainable parameters of a controller ANN scales at least quadratically with the number of beads $N$, the number of possible perceivable states and controlling actions scales in a combinatorial way with the number of degrees of freedom of the sensory input and the action output.
This not only immensely complicates deep-learning procedures \cite{Sutton1998} but essentially renders exhaustive approaches infeasible given the vast combinatorial space of possible input-output mappings for large $N$.
Thus, while generalized NG swimmers with $N\geq3$ have been successfully trained to perform locomotion or chemotaxis tasks \cite{Tsang2020,Hartl2021,Jebellat2024}, they have been limited to a relatively small number of body parts, \ie, a small number of degrees of freedom, $N\lesssim10$, so far.

However, even unicellular organisms are fundamentally made of (many) different parts, which (co-)operate in a seamlessly coordinated way:
in biological microswimmers, for example, collective large body deformations are typically achieved through orchestrated and cooperative action of molecular motors and other involved proteins, inducing, \eg,  the local deformation of the cilia-forming axoneme \textit{via} localized contractions and extensions \cite{Walczak1994,Bray2000}.
Consequently, such organisms -- without an apparent centralized controller -- cooperatively utilize their body components in a fully self-orchestrated way in order to swim by collectively deforming and reconfiguring their body shape.
Moreover, such \textit{decentralized} navigation policies tend to be robust and failure tolerant with respect to changing morphological or environmental conditions, e.g., if parts of the locomotive components are disabled or missing or unforeseeable situations are encountered.
Strong signs of such generalizing problem-solving skills are observed, for example, in cells \cite{Renkawitz2019}, slime molds \cite{Reid2012a}, and swarms \cite{Bonabeau1999,Kennedy2006}, and, as recently suggested \cite{McMillen2024}, this fundamental ability of biological systems to self-organize via collective decision-making might be the unifying organizational principle for integrating biology across scales and substrates.
Thus, the plastic and functional robustness and the innate drive for adaptability found in biological systems might not only further robotics \cite{Stoy2002, Kurokawa2006, Bongard2011, Pathak2019, Kriegman2020, Bing2020} but facilitate unconventional forms of computation based on collective intelligence \cite{Bongard2009, Zhu2018, Parsa2023}.

So far it remains unclear how decentralized decision-making in a deformable microswimmer can lead to efficient \textit{collective} locomotion of its body parts.
We thus investigate biologically motivated decentralized yet collective decision-making strategies of the swimming behavior of a generalized NG swimmer, which represents, \eg, a unicellular organism, or a controllable swimming microrobot.
Furthermore, in our approach, we are able to overcome earlier limitations by extending our swimmers to much larger $N$ than previously feasible, allowing us to identify locomotion strategies in the limit $N \rightarrow \infty$.
To this end, we interpret each bead of the microswimmer's body as an agent that can only perceive information about its adjacent beads and whose actions induce contractions or extensions of its adjacent muscles.
We substitute the internal decision-making machinery of such single-bead agents by ANNs and employ genetic algorithms and neuroevolution to machine learn optimal policies for such single-bead decision-making centers such that the entire $N$-bead swimmer can efficiently self-propel collectively, \ie, in a decentralized way.
We show that the evolved policies are robust and failure-tolerant concerning morphological changes of the collective microswimmer and that such decentralized control - trained for microswimmers with a specific number of beads - generalizes well to vastly different morphologies.

\section{System}
\label{sec:system}
\subsection{The N-Bead Swimmer Model}
\begin{figure*}
    \centering
    \includegraphics[width=\textwidth]{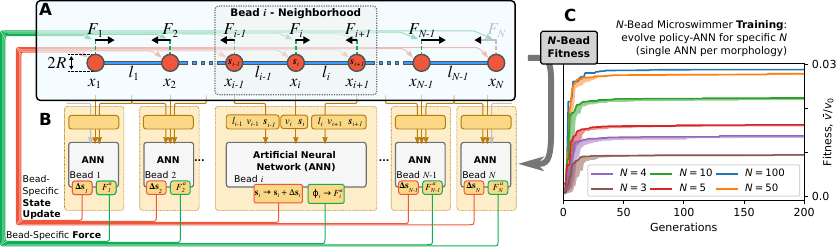}
    \caption{
    (\textbf{A}) Schematics of an $N$-bead microswimmer environment, with (\textbf{B}) functionally identical yet operationally independent Artificial Neural Networks (ANNs) acting as decentralized decision-making centers (or controllers) to update the respective internal states of the beads, $\mathbf{s}_i\rightarrow\mathbf{s}_i+\Delta\mathbf{s}_i$ (red arrows), and to apply bead-specific forces, $F_i \in [-2F_0,2F_0]$ (green arrows; ensuring $\sum_i F_i=0$), such that the entire microswimmer self-propels purely based on local perception-action cycles of the constituting bead controllers.
    (\textbf{C}) The training progress of optimizing various $N$-bead microswimmer locomotion policies of \typeA{} (see text and \cref{fig:system:type:AB}), respectively identifying for predefined values of \params{$(N=3-100)$} the parameters of the morphology-specific ANN controllers via evolutionary algorithms (EAs).
    The fitness score for different $N$, quantifying a specific $N$-bead center of mass velocity $\bar v$ (see \cref{sec:system,sec:methods}), is presented over \params{$200$} subsequent generations.
    Opaque-colored areas below the fitness trajectories indicate the corresponding STD of \params{$10$} independent EA searches per morphology and serve as a measure for convergence for the optimization process.
    }
        \label{fig:system}
\end{figure*}
Here, we investigate swimming strategies optimized by RL and the corresponding physical implications of $N$-bead generalized NG \cite{Najafi2004} swimmer models moving in a fluid of viscosity $\mu$. 
A swimmer consists of $N$  co-centrically aligned spheres of radius $R$ located at positions $x_i(t_k)$, $i=1\dotsc, N$, at time $t_k$.
These beads are connected pairwise by massless arms of length $l_i(t_k) = x_{i+1}(t_k) - x_{i}(t_k)$, as illustrated in \cref{fig:system}~(A).
The swimmer deforms and moves by applying time-dependent forces $F_i(t_k) = F^a_i(t_k) + F^r_i(t_k)$ on the beads.
The active forces $F^a_i(t_k)$ are proposed by RL agents (see below),
and passive restoring forces \cite{Hartl2021} $F^r_i(t_k)$ are applied when arm lengths $l_i(t_k)$ becomes smaller than $0.7L_0$ or lager than $1.3L_0$, where we choose $L_0=10R$ as the reference arm length.
The swimmer is force-free, $\sum_i F_i(t_k) =0$, and the bead velocities $v_i(t_k)$ are obtained in the Oseen approximation \cite{KimKarila}, $v_i =  F_i/(6\pi\mu R) + \sum_{j \ne i} F_j/(4 \pi \mu |x_i-x_j|)$, (see~\cref{sub:methods:ENV}).

\subsection*{Modeling system-level decision-making with decentralized controllers}
To identify the active forces $F^a_i(t_k)$ on the beads, we assign an ensemble of independent yet identical controllers to every bead which respectively can only perceive local information about adjacent beads (such as distances and velocities of their neighbors) and propose actions to update their respective states (such as proposing bead-specific forces to update their own positions).
Yet, these bead-specific agents follow a shared objective to collectively self-propel the entire $N$-bead swimmer's body.
More specifically - as illustrated in \cref{fig:system} and detailed in \cref{sub:methods:ANN} - for each time $t_k$ the controller associated with the bead $i$ perceives its left- and right-neighbor distances, $\mathcal{L}_i(t_k)=\{l_i(t_k), l_{i+1}(t_k)\}$, and its own- and the neighboring beads' velocities, $\mathcal{V}_i(t_k)=\{v_{i-1}(t_k), v_i(t_k), v_{i+1}(t_k)\}$.
Moreover, each bead maintains an internal vector-valued state, $\mathbf{s}_i(t_k)$.
This state can be utilized by every controller to store, update, and actively share recurrent information with other beads that is not necessarily bound to the physical state of the swimmer but an emergent property of the collective RL system:
Every controller thus perceives its neighboring states, $\mathcal{S}_i(t_i)=\{\mathbf{s}_{i-1}(t_k), \mathbf{s}_{i}(t_k), \mathbf{s}_{i+1}(t_k)\}$, which additionally guide the agent's decision-making.
In total, the perception of a single bead agent is given by $\mathbf{p}_i(t_k)=\{\mathcal{L}_i(t_k), \mathcal{V}_i(t_k), \mathcal{S}_i(t_k)\}$.

After integrating information about its local environment $\mathbf{p}_i(t_k)$, the controller of each bead $i$ computes, and then outputs an action, $\textbf{a}_i(t_k)=\{\phi_i(t_k), \Delta\mathbf{s}_i(t_k)\}$, comprising a \textit{proposed} active force, $\phi_i(t_k)$, and an internal state update, $\mathbf{s}_i(t_{k+1}) = \mathbf{s}_i(t_{k}) + \Delta{\mathbf{s}_i}(t_k)$ (see Fig.~\cref{fig:system}~B).
Notable, the proposed forces are limited to $\phi_i(t_k) \in [-F_0,F_0 ]$
where $F_0$ sets the force scale in our system and hence the maximum power consumption and bead velocities of the swimmer.

\begin{figure}[h]
    \centering
    \includegraphics[width=\columnwidth]{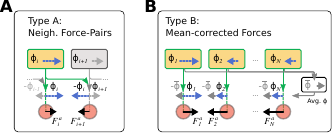}
    \caption{Schematics of mapping the bead-specific proposed actions $\phi_i(t_k)$ to the proposed active forces $F_i^a(t_k)$ to ensure the global force-free condition $\sum_{i=1}^N F_i^a(t_k)=0$ (see text) by either (\textbf{A}) interpreting the actions as force-pairs $F^a_i(t_k) = \phi_i(t_k) - \phi_{i-1}(t_k)$ between neighboring beads (\typeA{} microswimmers), or (\textbf{B}) subtracting the global average $\bar \phi(t_k)=\frac{1}{N}\sum_{i=1}^N\phi_i(t_k)$ from every proposed, bead-specific action $F^a_i(t_k)=\left(\phi_i(t_k)-\bar{\phi}(t_k)\right)$ (\typeB{} microswimmers).
    }
    \label{fig:system:type:AB}
\end{figure}

To model a force-free swimmer, we propose two different methods of how the mapping between the \textit{proposed} forces $\phi_i(t_k)$, and the \textit{actual} active forces $F^a_i(t_k)$, is achieved:
First, we interpret the proposed forces as pairwise arm forces $\phi_i(t_k)$ and $-\phi_i(t_k)$ applied between two consecutive beads $i$ and $i+1$, respectively (see \cref{fig:system:type:AB}~A).
This leads to the actual active forces $F^a_i(t_k) = \phi_i(t_k) - \phi_{i-1}(t_k)$ for beads $i=1\dotsc,N$, where we treat the swimmer's ``head'' and ``tail'' separately by setting $\phi_N(t_k)=0$ and introducing $\phi_0(t_k)=0$.
This automatically ensures $\sum_{i=1}^N F_i^a(t_k)=0$.
In this sense, the proposed actions can be understood as local decisions to expand/contract muscles between the beads where the maximum local power input on a bead is constrained.
Second, we assume that the proposed force $\phi_i(t_k)$ of every controller directly targets the actual force applied to its associated bead, but, to fulfill the force-free condition, we subtract the mean $\bar{\phi}(t_k)=\frac{1}{N}\sum_{j=1}^N\phi_j(t_k)$ from every proposed force and arrive at
$F^a_i(t_k)=\phi_i(t_k)-\bar{\phi}(t_k)$ (see \cref{fig:system:type:AB}~B).
Hence the first approach ensures the global force-free condition via a series of locally annihilating pair-forces motivated by biological force dipole generation at small scales that cause the arms between the corresponding beads $i$ and $(i+1)$ to contract or extend.
In turn, the second approach regularizes the forces by collective feedback (via $\bar{\phi}(t_k)$) and can be interpreted as a mean-field approach
that may be utilized by external controllers for artificial microswimmers \cite{Muinos-Landin2021}.
Henceforth, we refer to the first scenario as \textit{\typeA{}}, and to the second scenario as \textit{\typeB{}} microswimmers, and alike for the corresponding self-navigation strategies or policies.
We note that for both \typeAB{} microswimmers the total force per bead is constrained to $F_i(t_k) \in [-2F_0, 2F_0 ]$, except for the first and last bead of \typeA{} which are only connected to a single muscle such that for \typeA{} swimmers $F_0(t_k)$ and $F_N(t_k) \in [-F_0, F_0 ]$.

Following RL terminology, we refer to the mapping between perceptions and actions of an agent (or here synonymously, a controller) as its \textit{policy}, $\pi_i: \mathbf{p}_i(t_k)\rightarrow \mathbf{a}_i(t_k)$.
In general, such a policy is a complicated and complex function of the input, and Artificial Neural Networks (ANNs) as universal function approximators \cite{Hornik1989} are well-suited tools to parameterize these objects for arbitrary agents and environments (see \cref{sub:methods:ANN}).
Thus, we approximate the RL agent's policy $\pi_i$ by an ANN, formally expressed as a function $f_\theta(\cdot)$ with parameters $\theta$, such that $\mathbf{a}_i(t_k)=f_\theta(\mathbf{p}_i(t_k))$.
More specifically, we treat a single $N$-bead swimmer as a multi-agent system, each bead being equipped with a functionally identical but operationally independent ANN-based controller, $f_\theta(\cdot)$.
This renders the system reminiscent of a \textit{Neural Cellular Automaton} \cite{Mordvintsev2020} (NCA) with the extension that the decentralized actions of all individual controllers give rise to a collective locomotion policy $\Pi=\{\pi_1, \dotsc, \pi_N\}\approx\{f_\theta(\mathbf{p}_1(t_k)),\dotsc,f_\theta(\mathbf{p}_N(t_k))\}$, of the entire virtual organism (see also Ref.~\citenum{Risi2022a}).
Notably, only a single set of parameters $\theta$ is used for all $N$ bead-specific agents, \ie, the same ANN controller is deployed to every bead;
the states of the latter only differ in their initial conditions and subsequent input-output-history).
For our purposes, this renders the optimization problem much more tractable compared to situations with a single centralized controller, $\tilde\Pi\approx \tilde f_{\tilde\theta}(\mathbf{p}_1(t_k),\dotsc,\mathbf{p}_N(t_k))$, especially for large swimmer morphologies.

Here, we aim at identifying optimal and robust swimming gates for arbitrarily large $N$-bead swimmers, which translates to finding suitable ANN parameters, $\theta^\mathrm{(opt)}$, such that the bead-specific perception-action cycles, $\mathbf{a}_i(t_k)=f_{\theta^\mathrm{(opt)}}(\mathbf{p}_i(t_k))$, collectively self-propel the multi-agent system efficiently in a certain direction.
More specifically, the set of parameters $\theta$ comprises the weights and biases of the agent's ANN-based controller, which we designed to be numerically feasible for our neuroevolution approach (see below): 
In stark contrast to traditional RL agents, with often more than tens or hundreds of thousands of parameters, we here utilize a predefined architecture inspired by Refs.~\citenum{Hartl2024, Tang2021} with only \params{$59$} parameters (see \cref{fig:methods:ann}).
Thus, we can utilize evolutionary algorithms \cite{Katoch2020} (EAs), specifically a simple genetic algorithm \cite{Ha2017blog} discussed in \cref{sub:methods:EA}, to adapt the ANN parameters (but not the ANN topology \cite{Hartl2021}) such that the entire $N$-bead swimmer's mean center of mass (COM) velocity, $v_T=\frac{1}{N\,T}\left|\sum_{i=1}^N\left(x_i(T) - x_i(0) \right)\right|$, is maximized for a predefined swimming duration \params{$T=400-800 \Delta t$}, where $\Delta t=5\mu R^2/F_0$ is the time interval between two consecutive perception-action cycles of an RL agent.
$T$ is chosen sufficiently large to provide the respective $N$-bead swimmer enough time to approach a steady swimming state and to execute several swimming strokes, starting from randomized initial positions.
Thus, we define the objective, or fitness score as the mean COM velocity $r=\langle v_T\rangle_{N_e}$, averaged over $N_e=10$ statistically independent episodes, and search for $\theta^\mathrm{(opt)} = \max_{\delta\theta}(r)$ through variation $\delta \theta$ of the parameters $\theta$ via EAs, as detailed in \cref{sub:methods:EA}.

\FloatBarrier
\section{Results}
\label{sec:results}
\subsection*{Individual, bead-specific decisions facilitate collective swimming of an $N$-bead swimmer}
\label{sub:results:proof-of-concept}
We utilize EAs to optimize the parameters of the ANN-based controllers (which are deployed to every bead in a specific morphology) for different realizations of our multi-agent microswimmer models.
More specifically, we deploy morphologies ranging from $N=3$ to $N=100$ beads of \typeAB{} microswimmers and train every swimmer of different size $N$ for both types independently via EAs to self-propel by maximizing their respective fitness score $r$.
For details on the utilized ANNs and the applied EA we refer to \cref{sec:system,sec:methods}.

\begin{figure*}
    \centering
    \includegraphics[width=0.99\textwidth]{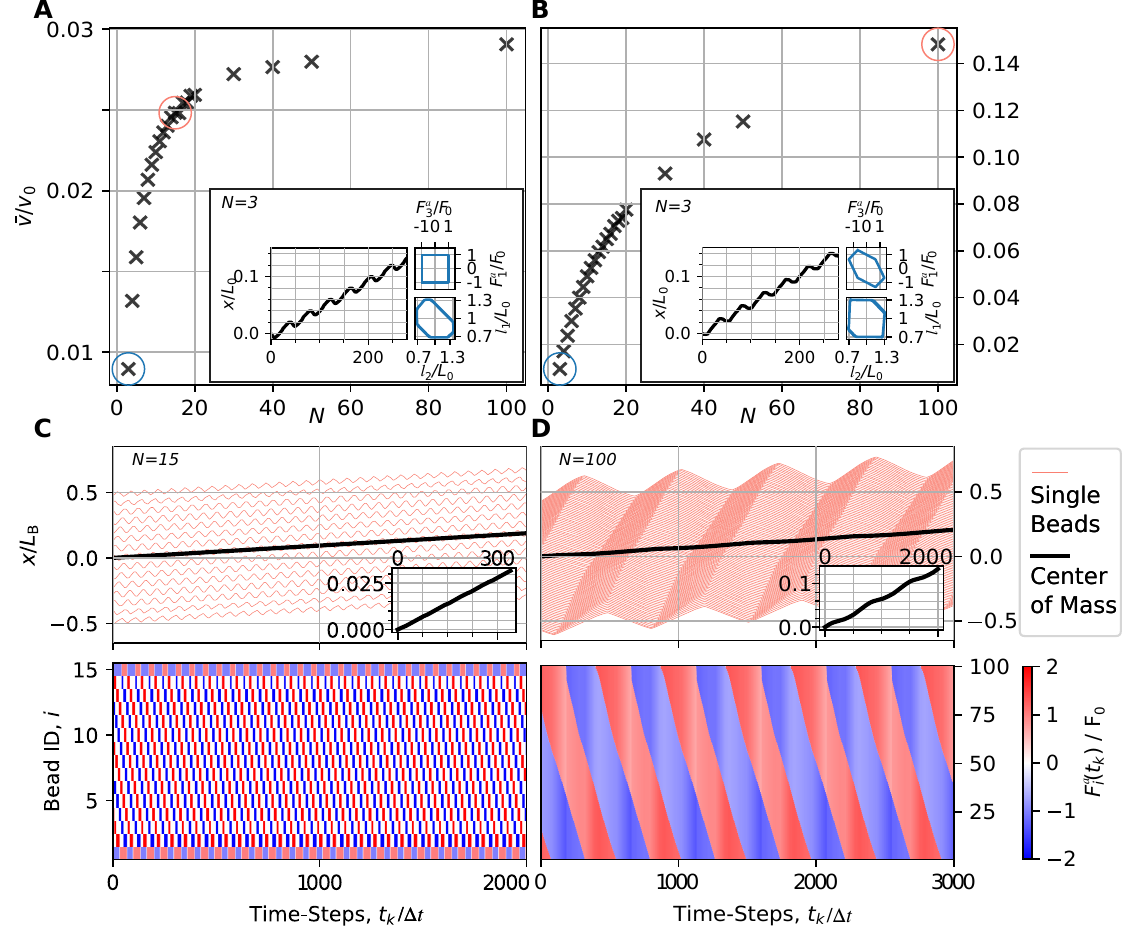}
    \caption{
        (\textbf{A}) and (\textbf{B}): Stroke-averaged COM velocity, $\bar v$, of different \typeAB{} microswimmers,
        respectively, corresponding to their fitness score when optimized independently with EAs for $N=3$ to $N=100$ beads (see also \cref{fig:system}~(C)).
        Insets show COM trajectories, and $(F^a_1, F^a_3)$- and $(l_1, l_2)$-phase-space plots for  $N=3$ (see blue circles).
        (\textbf{C}) and (\textbf{D}): Typical bead-specific coordinate- (top panels) and force-trajectories (bottom panels) of an \params{$(N=15)$-bead} \typeA{} and an \params{$(N=100)$}-bead \typeB{} microswimmer, respectively (the examples, see red circles in (\textbf{A, B}), are chosen for illustrative purposes and are representative for all investigated $N$ for both \typeAB{} policies). See also Movies S1-3.
        Insets detail the corresponding COM trajectories.
        For \typeA{} microswimmers (\textbf{A, C}), periodic localized waves of arm strokes travel through the body.
        In contrast \textbf{(B, D)}, large-scale collective body contractions allow large \typeB{} microswimmers to propagate much faster.
        Coordinate trajectories are normalized by the reference arm length $L_0$ in the insets of (\textbf{A,B}), and by the total sum of the reference arm lengths $L_\mathrm{B}=L_0\times(N-1)$ in (\textbf{C,D}).
        }
    \label{fig:results:collective-policy}
\end{figure*}

The training progress is presented in \cref{fig:system}~(C) exemplarily for \typeA{} microswimmers of different morphologies $N$, demonstrating that the proposed decentralized decision-making strategy is capable of facilitating fast system-level swimming gates for all the considered swimmer sizes up to $N=100$.
Thus, our method removes the bottleneck for machine-learning navigation policies of large-scale microswimmers by employing computationally manageable local ANN-based perception-action loops of their bead-specific agents.
To the best of our knowledge, this is the first time successful training of microswimmers with such a large number of degrees of freedom has been achieved.

\subsection*{Different strategies of autonomy: Large-scale coordination enables fast swimming}
\label{sub:results:collective-policy}
Employing the learned policies of both \typeAB{} microswimmers for different body sizes, \ie the number of beads $N$, we determine the respective stroke-averaged COM velocities $\bar{v}$,
which increase monotonously with $N$ as depicted in \cref{fig:results:collective-policy}~(A,B).
We normalize all velocities here with $v_0=2F_0/(6\pi\mu R)$, \ie\ the velocity of a bead dragged by an external force of strength $2F_0$.
Interestingly, \typeB{} swimmers are significantly faster compared to \typeA{} swimmers by almost one order of magnitude, especially for large $N$.
As illustrated in \cref{fig:results:collective-policy}~(A), for \typeA{} microswimmers with locally ensured force-free conditions, the mean COM velocity $\bar v$ saturates with increasing $N$ 
at \params{$\bar{v}_\mathrm{max}/v_0\approx0.03$ for $N=100$}.
In contrast (c.f., \cref{fig:results:collective-policy}~(B)), the fastest \typeB{} microswimmer, again at $N=100$, achieves a maximum COM velocity of \params{$\bar{v}_\mathrm{max}/v_0\approx0.15$}.

The insets in panels (A, B) illustrate results for the well-studied three-bead swimmer ($N=3$).
First, they show the characteristic periodic \textit{1-step-back-2-step-forward} motion of the 
COM trajectory \cite{Najafi2004}.
Second, the corresponding steady state phase space dynamics of the active forces on beads 1 and 3, $(F^a_1, F^a_3)$, and of the arm lengths $(l_1, l_2)$, reiterating the periodic motion.
Note that the force on bead 2 simply follows from $F^a_2 = - F^a_1 - F^a_3$.
While both \typeAB{} three-bead swimmers move at comparable speed, this is achieved with different policies, as can be seen by the different phase space curves (see also Movie S1).

In panels (C) and (D) we present trajectories of both the COM and bead-specific coordinates (top panels; respective insets emphasize the COM dynamics), and of the bead-specific proposed forces (bottom panels) for an $(N=15)$-bead \typeA{}- and $(N=100)$-bead \typeB{} microswimmer, respectively.
These selected trajectories demonstrate the genuinely different swimming strategies for \typeAB{} microswimmers (which have notably been optimized with the same EA settings).

In \typeA{} microswimmers (\cref{fig:results:collective-policy}~(C)), the pairwise arm forces induce periodic waves of arm contractions of relatively high frequency but small wavelength, which travel through- and move forward the body of the $N$-bead microswimmer.
For swimmers with a sufficiently large number of beads $N$ this leads to a relatively smooth and linear COM motion (see inset in \cref{fig:results:collective-policy}~(C)).

In stark contrast, the fastest swimming strategies for \typeB{} microswimmers (\cref{fig:results:collective-policy}~(D)) assumes coordinated arm strokes across large fractions of their bodies, essentially contracting and extending almost the entire swimmer simultaneously, which is reflected in the oscillatory COM motion even for very large $N$ (see inset in \cref{fig:results:collective-policy}~(D)).
This large-scale coordination exceeds the capabilities of the locally interlocked policies of \typeA{} microswimmers and strikes us as an emergent phenomenon \cite{Anderson1972} which - still based on purely local decision-making - is facilitated by the mean-field motivated feedback of the mean proposed force of all the agents in the system:
\typeB{} microswimmers seemingly act as a single entity \cite{Levin2019}, despite the fact that the precise number of constituents, $N$, is not important and can vary (see also Movies S2 and S3).

As shown in \cref{fig:results:collective-policy}~(D), typical emergent swimming gaits of the respective \typeB{} swimmers are reminiscent of the large amplitude contraction-wave based locomotion of crawling animals such as caterpillars \cite{VanGriethuijsen2014}.
Similarly, the crawling locomotion of crawling fly larvae had been optimized using RL recently \cite{Mishra2020a}.
In the context of locomotion in viscous fluids, large-amplitude traveling waves along the body have been discussed as an alternative swimming strategy of flagella-less organisms \cite{Ehlers1996a,Najafi2005}.

\subsection*{Transferable evolved policies: decentralized decision-making generalizes to arbitrary morphologies}
\label{sub:results:generalize}

\begin{SCfigure*}
    \centering
    \includegraphics[width=11.4cm]{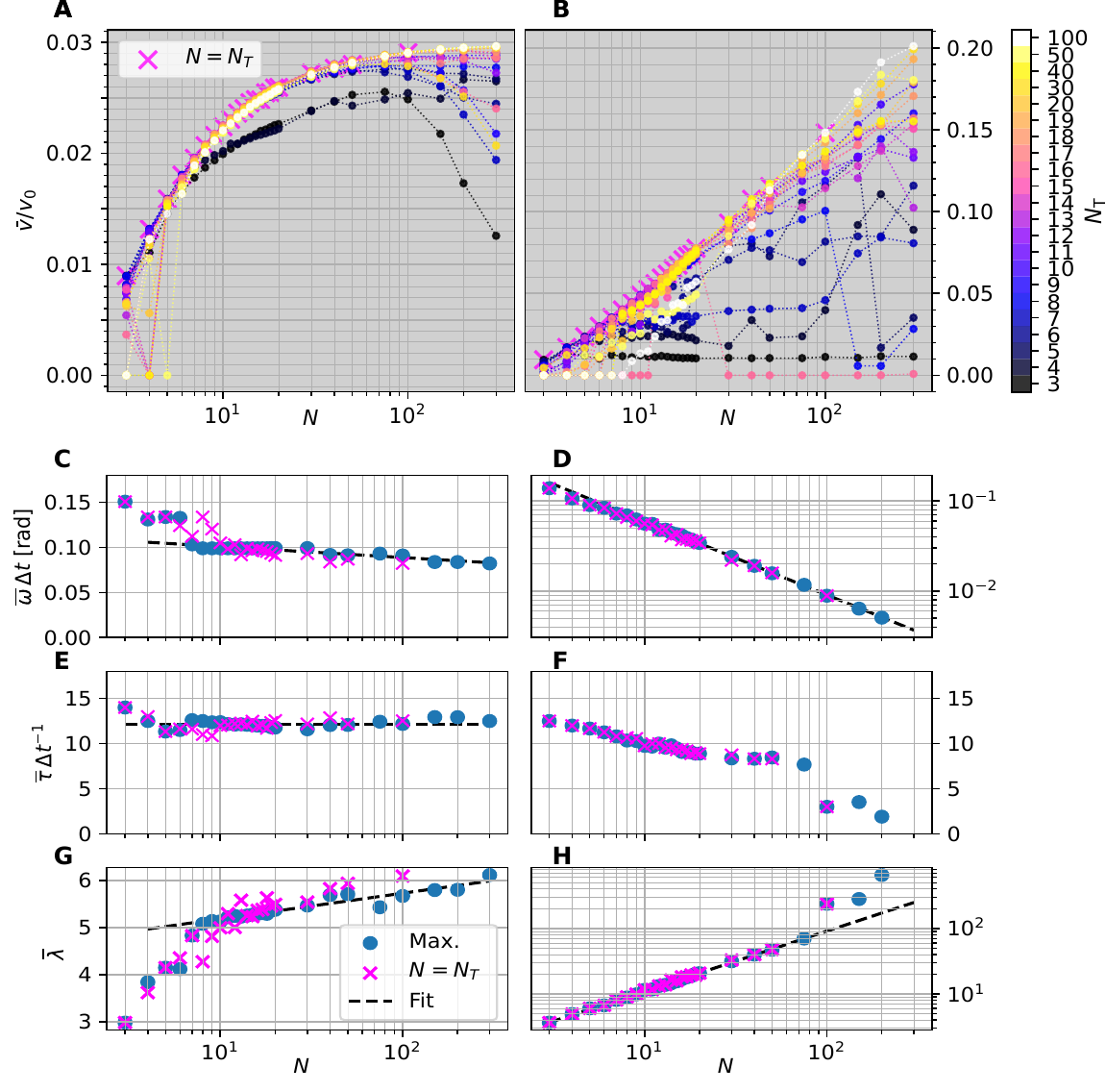}
    \caption{
        (\textbf{A}) and (\textbf{B}): Cross-policy \textit{transferability} evaluations, depicting the stroke-averaged COM velocity $\bar{v}$ of $N$-bead microswimmers deployed with policies optimized for $N_\mathrm{T}$ (color-coding) for \typeAB{} microswimmers, respectively.
        The vast majority of the policies evolved for $N_\mathrm{T}$-beads generalize well to vastly different $N$-bead morphologies without further optimization.
        (\textbf{C}) and (\textbf{D}): The mean  angular velocity $\bar\omega$ of the arm-length limit-cycle dynamics of the most optimal $N$-bead \typeAB{} microswimmers, respectively (blue circles; fastest policies trained with $N_\mathrm{T}$-beads and deployed to $N$-bead microswimmers);
        magenta $\times$ symbols illustrate $\bar\omega$ for the corresponding training conditions, $N=N_\mathrm{T}$.
        (\textbf{E,F}) and (\textbf{{G,H}}): Similar to (\textbf{C,D}), but showing the mean cross-correlation time $\bar\tau$ ((\textbf{E,F}), see text) between neighboring arm lengths,
        and the corresponding (dimensionless) wavelength $\bar{\lambda}$ ((\textbf{G,H}), see columns in lower panels of \cref{fig:results:collective-policy}~(C, D) at fixed time $t_k$ and see text) of \typeAB{} microswimmers, respectively, as a function of $N$.
        Dashed lines indicate functional fits: 
        (\textbf{C}) $\bar\omega\approx a \ln N + b$ with $a = -5.26\times10^{-3}~\textrm{rad}/\Delta t$ and $b = 1.13\times10^{-1}~\textrm{rad}/\Delta t$.
        (\textbf{D}) $\bar\omega\approx\alpha N^\beta$ with 
        $\alpha=3.92\times10^{-1}~\textnormal{rad}/\Delta t$ and $\beta = -8.19\times10^{-1}$. 
        (\textbf{E}) $\bar\tau\approx12.2~\Delta t$. 
        (\textbf{G}) $\bar\lambda\approx a \ln N + b$ with $a = 2.37\times10^{-1}$ and $b = 4.64$. 
        (\textbf{H}) $\bar\lambda\approx\alpha N^\beta$  with $\alpha=1.33$ and $\beta = 9.17\times10^{-1}$.
    }
    \label{fig:results:generalize}
\end{SCfigure*}

We have recently shown that biologically inspired NCA-based multi-agent policies - especially when evolved via evolutionary processes - can display behavior that is highly robust against structural and functional perturbations and exhibit increased generalizability, adaptability, and transferability \cite{Hartl2024, Manicka2022}.
We thus investigate here, whether our decentralized locomotion policies, which are genuinely evolved for microswimmer bodies with exactly $N_\mathrm{T}$ beads, generalize to morphological changes.
More specifically, we carefully optimize ANN policies for a particular number of $N_\mathrm{T}=3$ to~$100$ beads (as discussed above) and deploy them - without any retraining or further adaptation - into microswimmer bodies with a different number of $N=3$ to~$300$ beads instead to evaluate the corresponding swimming velocities for such \textit{cross-policy} environments.

Remarkably, as illustrated in \cref{fig:results:generalize}(A) and (B), we find that the vast majority of all policies that are most optimal for a particular value of $N_\mathrm{T}$ are also highly effective in self-propelling microswimmers with $N\neq N_\mathrm{T}$ for both \typeAB{}, respectively.
This even holds for situations where $N_\mathrm{T}\ll N$, such as $N_\mathrm{T}=3$ and $N=300$.
Notably, we did not explicitly optimize for this property at all, but it is an emergent phenomenon of the inherent decentralized decision-making of the system.
Thus, the collective nature of the proposed swimming policies renders the evolved locomotive strategies highly adaptive to morphological changes, irrespective of the specific number of beads $N$ used during deployment.
Only a few combinations of $N_T$ and $N$, in particular for \typeB{} microswimmers, are unsuccessful and do not lead to (fast) locomotion.

Moreover, the set of policies evolved and deployed for different bead numbers of $N_\mathrm{T}$ and $N$, respectively, allows us to estimate a trend for optimal swimming gates for large $N$:
In steady state, the arm-lengths $l_i(t_k)$ of an arbitrary trained $N_\mathrm{T}$-bead swimmer describe limit cycle dynamics in an $(N-1)$-dimensional phase space (c.f, \cref{fig:results:collective-policy}).
Then, all arms oscillate at the same angular velocity $\bar\omega$ and exhibit the same cross-correlation times $\bar\tau$ to their neighboring arms, where $l_{i+1}(t_k) = l_{i}(t_k-\bar\tau)$, and we can express each of the $i=1,\dotsc,N$ arm lengths as a $2\pi$-periodic function $\hat l_i(t) = f_\ell\left((t - i\,\bar\tau)\,\bar\omega + \phi\right)$, with the phase shift $\phi$ defining the initial conditions; 
for more details about $\bar\omega$ and $\bar\tau$ we refer to \cref{sub:methods:FFT}.
Notably, we can also write $\hat l_i(t)$ in the form of a wave-equation as
$\hat l_i(t) = f_\ell\left(t\bar\omega +  2\pi i/\bar{\lambda} + \phi\right)$, where the bead index $i$ controls the ``spatial'' oscillations at a (dimensionless) wavelength $\bar{\lambda}=\frac{2\pi}{\bar\omega\bar\tau}$ irrespective of the corresponding physical bead positions $x_i(t_k)$.

In \cref{fig:results:generalize}~(C, E, G), we respectively present $\bar\omega$, $\bar\tau$, and $\bar{\lambda}$ for the fastest \typeA{} microswimmers as a function of $N$ (blue circles), additionally to the training conditions where $N=N_\mathrm{T}$ (magenta ``$\times$'' symbol).
We can see that these quantities are almost independent of $N$, and we observe only weak logarithmic dependencies of $\bar\tau$ and $\bar{\lambda}$ (see \cref{fig:results:generalize} caption).
In contrast, for \typeB{} (see \cref{fig:results:generalize}~(D, F, H)), the angular velocity is approximately inverse to the swimmer length $\bar\omega \sim N^{-1}$, and the wavelength almost linear to $N$, $ \bar\lambda \sim N$ (detailed fits see \cref{fig:results:generalize} caption).
As a result, the evaluated $\bar\tau$ values for \typeB{} microswimmers (\cref{fig:results:generalize}~(F))
are almost constant and the values are comparable to those of the \typeA{} microswimmer, but slightly decrease with $N$.

\subsection*{Large-scale coordination leads to efficient locomotion}
\label{sub:results:efficiency}

\begin{SCfigure*}
    \centering
    \includegraphics[width=10.3cm]{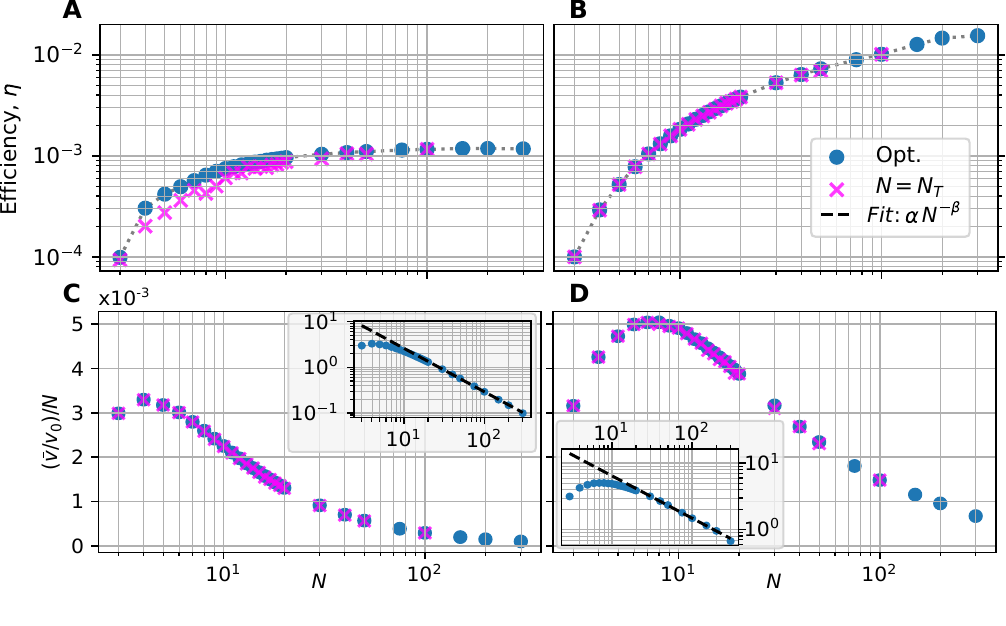}
    \caption{
    (\textbf{A}) and (\textbf{B}): Hydrodynamic efficiency $\eta$ of the most optimal $N$-bead \typeAB{} microswimmers (blue circles), respectively;
    magenta $\times$ symbols emphasize the efficiency of microswimmers trained with $N=N_\mathrm{T}$.  
    (\textbf{C}) and (\textbf{D}): 
    stroke-averaged COM velocity per bead $\bar v/N$ for the  respective \typeAB{} microswimmers.
    The insets show a power law decay, $(\bar v/v_0)/N=\alpha\,N^{-\beta}$, with 
    $\alpha=2.34\times10^{-2}$, $\beta=9.53\times10^{-1}$ (\textbf{C}) and 
    $\alpha=2.83\times10^{-2}$, $\beta=6.41\times10^{-1}$ (\textbf{D}), (black dashed lines).
    }
    \label{fig:results:efficiency}
\end{SCfigure*}

In our approach we limit the maximum forces on each of the beads, and thus the maximum bead velocities. This procedure is somewhat similar as fixing the total power consumption of the swimmer.
In previous optimization procedures on 3- and N-bead swimmers commonly the swimming efficiency is optimized \cite{Nasouri2019,Wang2019c}, where e.g.\ the power consumption of the entire swimmer is taken into account as a single, global quantity.
In contrast, in our work we set local constraints on the swimmer by limiting the forces on every bead.
Although we hence did not optimize in our RL procedure for the hydrodynamic efficiency $\eta$ of our swimmers, we measure $\eta$ determined by \cite{Lauga2009a,Nasouri2019} $\eta = 6\pi\mu R_{\mathrm{eff}}\bar{v}^2/\mathcal{P}$ where $\mathcal{P}=\frac{1}{T}\int \sum_i v_i(t) F_i(t)\mathrm{d}t$ is the stroke-averaged power consumption, and $R_{\mathrm{eff}}$ is the effective radius of our swimmers.
There is no unique way to define $R_{\mathrm{eff}}$ of our swimmer (see also the discussion in Ref. \cite{Nasouri2019}), and it is hence not straightforward to compare $R_{\mathrm{eff}}$ of swimmers of different size $N$.
We choose here $R_{\mathrm{eff}}=NR$ which approximates the combined drag on all of the spheres neglecting hydrodynamic interactions.
The power consumption is naturally limited to be $\mathcal{P}<\mathcal{P}_\mathrm{max}$ with $\mathcal{P}_\mathrm{max} = 2NF_0v_0=2NF_0^2/(6\pi\mu R)$, for both cases \typeAB{}.
However \typeB{} swimmers can exploit their higher freedom to adjust the forces on the beads compared to the arm-force-limited \typeA{} swimmer to locomote at higher speed, and hence at higher efficiency $\eta$.
As seen in \cref{fig:results:efficiency}~(A,B) for both \typeAB{} swimmers the efficiency increases with swimmer length $N$ and levels off at large $N$.
The efficiency is relatively low for all swimmer lengths $N$ for \typeA{} swimmers, and is limited to $\approx 0.12\%$. 
In contrast, long \typeB{} swimmers can reach surprisingly high efficiencies of even $\approx 1.5\%$ for $N=100$, comparable to the efficiency of real existing microswimmers \cite{Lauga2009a}.

As discussed above, a larger microswimmer can swim faster due to the emergence of long-wavelength longitudinal waves.
Indeed it thus is not surprising that longer \typeB{} swimmers are faster than shorter ones.
Animals typically scale their speed with their size \cite{Meyer-Vernet2015}.
Here we determine the swimmer speed per size $\bar{v}/N$ depending on $N$ as shown in \cref{fig:results:efficiency}~(C,D), where we identify a maximum for $N=4$ for \typeA{} and at $N=8$ for \typeB{}.
Hence these swimmers need the smallest amount of time to swim their own body size $NL_0$ compared to swimmers of different $N$.
For large $N$, $\bar{v}/N$ decays with a power law (see inset \cref{fig:results:efficiency}~(C,D)).

\subsection*{Towards drug delivery: Robust- and failure tolerant locomotion allows cargo transport without re-training}
\label{sub:results:cargo}

\begin{figure*}
    \centering
    \includegraphics[width=\textwidth]{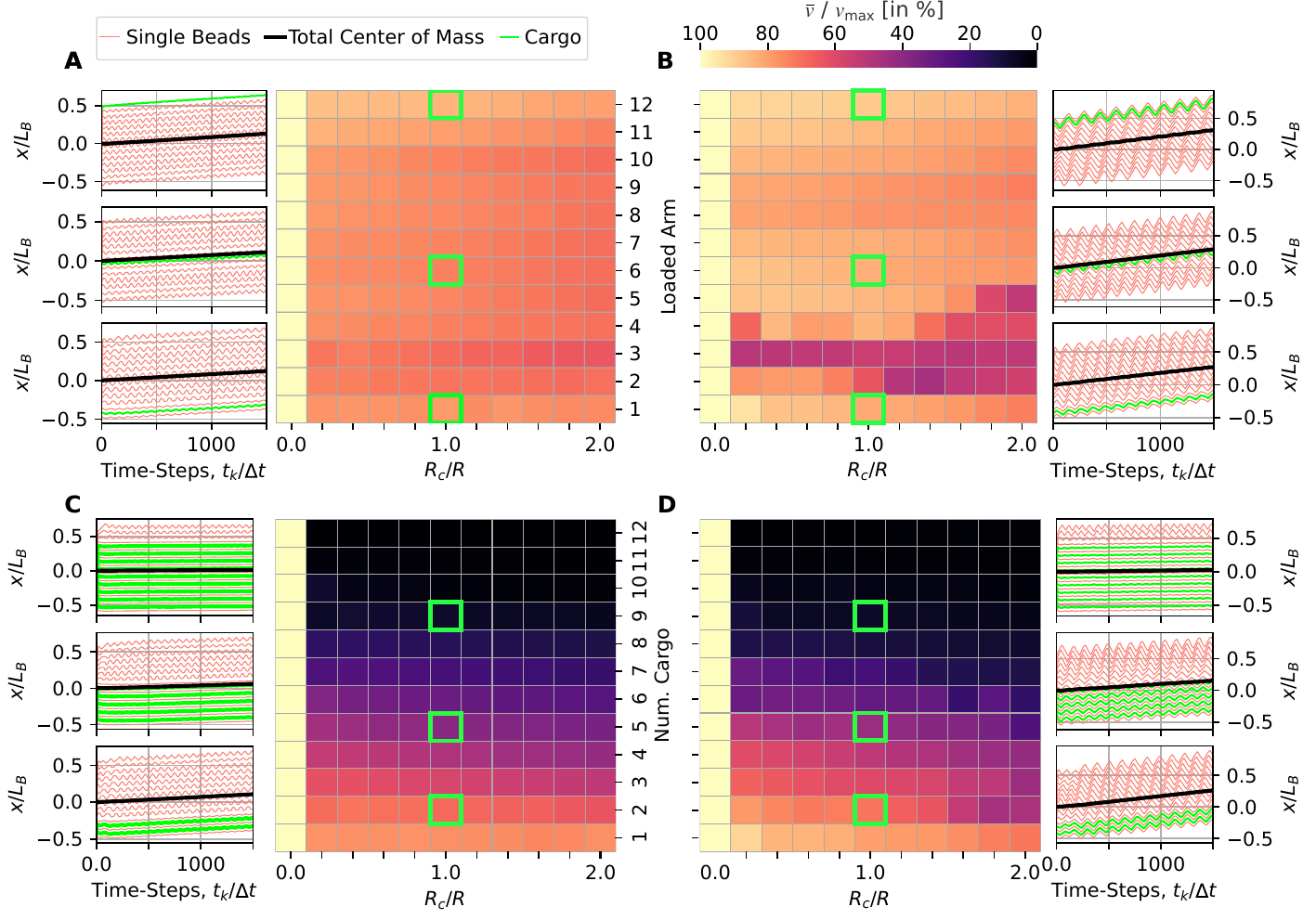}
    \caption{
    (\textbf{A}) and (\textbf{B}): Stroke-averaged COM velocity $\bar v$ in \% (color-coded) of the maximum velocity $v_\mathrm{max}$ (c.f., respective left columns) of representative $(N=13)$-bead \typeAB{} microswimmer solutions, respectively, carrying a cargo bead of radius $R_c\in[0, 2R]$ (horizontal axis) at different loading positions, \ie, at the respective arms $l_1$ through $l_{N-1}$ (vertical axis).
    For each panel, three selected trajectory plots (indicated by green frames in respective heatmaps) highlight the corresponding cargo-swimmers' single-bead- (red), COM- (black), and cargo bead trajectories (green) scaled by the total length of all arms $L_\mathrm{B}=(N-1)L_0$.
    (\textbf{C}) and (\textbf{D}): Similar to (\textbf{A}) and (\textbf{B}) but loading the respective $(N=13$)-bead microswimmers with a total number of $N_c=1,\dotsc, (N-1)$ cargo beads (vertical axis) of the same size, $R_c$ (horizontal axis), where each cargo occupies a single arm $l_i$  starting from $i=1,\dotsc,N_c$ (c.f., green cargo lines in designated example trajectories).
    Note, for the reference case, $R_c=0$, we do not include any cargo beads in the simulations, thus the corresponding velocity represents $v_\mathrm{max}$ for \typeAB{}, respectively.
    }
    \label{fig:results:cargo}
\end{figure*}
Since our evolved microswimmer policies show strong resilience even against significant morphological perturbations (see \cref{fig:results:generalize}), we aim to investigate this operational plasticity even further:
Exemplarily for the most optimal \typeAB{} microswimmers with \params{$(N=13)$}, we systematically load the arms with extra passive ``cargo'' beads of variable radius $R_c\in[0, 2R]$, and evaluate the dynamics of the corresponding loaded microswimmer consisting now of $N=13+1$ beads, without retraining or further optimization.

These cargo beads can geometrically be located between two neighboring beads of the microswimmer, but remain functionally disjoint from the latter (cargo beads are not connected by arms to any body beads): 
when placed at an arm $l_i$, a cargo bead does not disrupt the exchange of information between the corresponding beads $i$ and $(i+1)$ and thus does not affect the inputs of the respective bead-specific ANNs.
Since cargo beads are passive elements, they do not propose active forces independently, $F_c^a(t_k)=0$, and are moved around solely by the hydrodynamic interaction with the other beads and the restoring spring forces of nearby beads.
This ensures that a cargo bead is topologically fixed in the microswimmer's body.

\Cref{fig:results:cargo}~(A) and (B) demonstrates, that our approach of utilizing decentralized, bead-specific controllers for $N$-bead microswimmer locomotion not only gives rise to highly robust self-propulsion policies (see \cref{sub:results:generalize}) but can also be used ``out of the box'' for cargo transport applications \cite{Abdallah2020}:
We show that both \typeAB{} microswimmers are capable of swimming while having additional cargo beads of various sizes located at different internal positions, \ie, at different arms along their main axes.
While the presence of cargo beads significantly restrain the neighboring beads from adapting the adjacent arm-length, essentially locking the corresponding arm, the remaining functional beads self-propel the entire microswimmer effectively.
We further emphasize that, in general, the swimming speed decreases with increasing cargo size, $R_c$.

Next, we successively fill all arms of a single microswimmer from the left, $l_1$, to the right, $l_{N_c}$, simultaneously with a total number of $N_c=1,\dotsc,(N-1)$ cargo beads of equal radius $R_c$ (one cargo per arm) and measure the speed of the correspondingly loaded $(N+N_c)$-bead microswimmers as a function of the number of loaded arms and cargo size.
Both \typeAB{} microswimmers are capable of transporting multiple cargo loads at once efficiently, as illustrated in \Cref{fig:results:cargo}~(C) and (D):
they can carry up to \params{$\approx50-60\%$} of their active beads $N$ before the respective locomotion capabilities fail due to an increasing number of blocked arms.

Thus, our evolved navigation policies of the proposed microswimmer system not only show strong resilience against partly significant morphological changes of the swimmer's body (see also \cref{sub:results:generalize}), but even against functional failures of single or multiple actuators (c.f., blocked arms in \cref{fig:results:cargo}).
We emphasize that this goes well beyond what the swimmers experienced during training and is a clear sign of generalizability.
Notably, while we fixed $N=13$ to demonstrate cargo transport, it can be applied to different swimmer realizations, \ie, other morphologies and evolved policies.

\section{Conclusion}
\label{sec:conclusion}

Our study demonstrates that machine-learning of decentralized decision-making strategies of the distributed actuators in a composite \insilico microswimmer can lead to highly efficient navigation policies of the entire organism that are, moreover, highly robust with respect to morphological changes or defects.
More specifically, we treat each of the $N$ beads of a generalized NG microswimmer model as an ANN-based agent that perceives information about its adjacent beads and whose actions induce activations of adjacent muscles.
Via genetic algorithms, we have optimized such single-bead decision-making centers to collectively facilitate highly efficient swimming gates on the system level of the NG swimmer.

In that way, we have identified locomotion strategies for increasingly large microswimmer bodies, ranging from $N=3$ to $N=100$, with hydrodynamic efficiencies of up to $\eta\approx1.5\%$, close to that of real biological microswimmers;
to the best of our knowledge, this is the first demonstration of successfully training an $(N=100)$-bead microswimmer.

While having focused here on evolving swimming gates for NG microswimmers of fixed morphologies, we report that the optimized decentralized locomotion policies generalize well without any further optimization towards partly severe morphological changes:
policies optimized for an $N_\mathrm{T}$-bead microswimmer are in most cases also highly effective for ${(N\neq N_\mathrm{T})}$-bead morphologies, even if $N\gg N_\mathrm{T}$ or vice versa.
This renders our approach robust and  modular from both a physiological and information processing point of view \cite{Risi2022a,Hartl2024}, going well beyond more traditional, in this sense much more brittle RL applications that are typically concerned with centralized controllers with a fixed number of inputs and outputs.

The limiting computational factor in our simulations is not the controller part, but the $\mathcal{O}(N^2)$ complexity of the hydrodynamic model, which could be leveraged by further modeling, numerical optimization, or hardware accelerators.
However, the scalability of our approach allows us to generalize the optimized policies as a function of $N$, again overcoming limitations posed by traditional RL methods, and leveraging analytical investigations \cite{Wang2019c} of the generalized NG model to the limit of $N\rightarrow\infty$.

Intriguingly, and as we demonstrate, the inherent robustness and increased structural and functional plasticity of the here investigated microswimmer locomotion policies, based on decentralized (collective) decision-making, makes our system directly suitable for cargo transport applications without further optimization or fine-tuning.
Since the here-proposed distributed ANN controllers and learning paradigm are not limited to integrating information of the local neighborhood of a single bead in an $N$-bead NG microswimmer, our approach can be extended to virtually arbitrary swimmer geometries and models.
Thus, our approach represents a promising framework to develop autonomous cargo transport- \cite{Abdallah2020}
or biomedically relevant drug-delivery systems \cite{Jang2019, Singh2019, Patra2013, Kievit2011}, especially when combined with chemotactic capabilities \cite{Hartl2021}.

Our interdisciplinary approach, integrating cutting-edge concepts from biology, biophysics, robotics, collective artificial intelligence, and artificial life \cite{Langton1997}, offers a promising path to designing and understanding robust and fault-tolerant microswimmer policies that are computationally efficient.
Reminiscent to the remarkable structural and functional plasticity of ``real'' biological matter \cite{Cooke1981}, we emphasize the striking inherent ability of our microswimmer policies to adapt without any retraining to functional perturbations or morphological changes ``out of the box''.
This resonates well with William James' definition of intelligence \cite{FieldsLevin2023, McMillen2024} of ``achieving a fixed goal with variable means'', and raises interesting philosophical questions about the nature of emergent individuality \cite{Watson2022} and the role of collective intelligence \cite{McMillen2024} in multi-agent systems inspired by the multi-scale competency architecture of biology \cite{Hartl2024, Levin2023, Levin2022TAME}.


\section{Methods}
\label{sec:methods}
\FloatBarrier
\subsection{Hydrodynamic interactions and numerical details for the $N$-bead swimmer model}
\label{sub:methods:ENV}
The microswimmer consists of $N$ hydrodynamically interacting beads located at positions $x_i(t_k)$, $i=1,\dots,N$, at time $t_k$.
The bead positions change over time by applying forces $F_i(t_k)$, consisting of active ($F^a_i(t_k)$) and passive ($F^r_i(t_k)$) contributions.
At time $t_k$ the velocities of the beads $v_i(t_k)$ depend linearly on the applied forces through the mobility tensor $\mathcal{M}(t_k)$: $v_i(t_k) = \sum_j\mathcal{M}_{ij}(t_k)F_j(t_k)$.
Self-mobilities are given by Stokes formula $\mathcal{M}_{ii} = 1/(6\pi \mu R)$,
while cross-mobilities describe hydrodynamic interactions, which we consider in the far-field limit in the Oseen approximation: $\mathcal{M}_{ij}(t_k) = 1/(4 \pi \mu |x_i(t_k) - x_j(t_k)|)$.
Active forces $F_i^a(t_k)$ are applied as described in the main text.
We apply passive harmonic spring forces as pairwise restoring forces $F_{i,i+1}^r(t_k)$ between beads $i$ and $i+1$.
They depend on the arm length $l_i(t_k)$ between the beads and are applied if $l_i(t_k)<0.7L_0$ such that $F_{i,i+1}^r(t_k) = k (l_i(t_k) - 0.7L_0) < 0$,
or if $l_i(t_k) > 1.3L_0$ such that $F_{i,i+1}^r(t_k) = k (l_i(t_k) - 1.3L_0) > 0$, where $k=10F_0/R$ is the spring constant.
Every bead is potentially affected by restoring forces between both neighbor beads,
resulting in a total restoring force
$F_i^r(t_k) = -F_{i-1,i}^r(t_k) + F_{i,i+1}^r(t_k)$,
except for the beads at the end which only have a single neighbor bead 
where $F_1^r(t_k) = F_{1,2}^r(t_k)$
and $F_N^r(t_k) = -F_{N-1,N}^r(t_k)$.
This procedure limits the arm extensions, as can be seen for example in the inset of \cref{fig:results:collective-policy}~(A,B).
Note, both the active and passive forces sum up to zero individually, $\sum_i F_i^r(t_k) = 0$ and $\sum_i F_i^a(t_k)$.
The equations of motion of the microswimmer are then solved using a fourth-order Runge Kutta scheme using a sufficiently small time step $\delta t = \Delta t/10$. 

\FloatBarrier
\subsection{Artificial Neural Network-based Decentralized Controllers}
\label{sub:methods:ANN}
Mimicking the flexible operations of biological neural circuits, \textit{Artificial Neural Networks} (ANNs) consisting of interconnected \textit{Artificial Neurons} (ANs) have become invaluable numerical tools for statistical learning applications \cite{Goodfellow2016}.
Each AN takes a set of inputs,  $\mathbf{x}\in\mathbb{R}^n$, and maps them onto a single output value, $y\in\mathbb{R}$, through a weighted non-linear filter, $y=\sigma(\mathbf{w}\cdot\mathbf{x}+b)$, where the weights $\mathbf{w}\in\mathbb{R}^n$ represent the strengths of every individual input connection, and $b\in\mathbb{R}$ is the bias, representing the AN's firing threshold \cite{Minsky1969}.

ANNs are commonly organized into layers of ANs. 
A \textit{Feed Forward} (FF) ANN transforms an input, $\mathbf{x}^{(1)}\in\mathbb{R}^\mathrm{N_0}$, through a series of hidden layers ($i=1,\dotsc, N_\mathrm{L}$) to an output vector, $\mathbf{y}^\textnormal{(out)}\in\mathbb{R}^{N_\mathrm{L}}$. 
Each layer's output is calculated as $\mathbf{y}^{(i)}=\sigma\left(\mathcal{W}^{(i)}\cdot\mathbf{x}^{(i)} + \mathbf{b}^{(i)}\right)$, where $\mathcal{W}^{(i)}=\{w^{(i)}_{jk}\}\in\mathbb{R}^\mathrm{N_{i}\times N_{i-1}}$ is the weight matrix and $\mathbf{b}^{(i)}\in\mathbb{R}^\mathrm{N_{i}}$ is the bias vector. 
In a FF ANN, the output of layer $i$ becomes the input to the next deeper layer $(i+1)$ through successive dot-products, until an output is generated.
Training an ANN thus involves optimizing a set of parameters, $\theta=\{w_{jk}^{(i)}, b^{(i)}_k\}$, \ie, the entire network's weights and biases, such that the ANN's response to known inputs have minimal deviation to (typically predefined) desired outputs \cite{Rumelhart1986, LeCun2015}.

\begin{figure}
    \centering
    \includegraphics[width=\columnwidth]{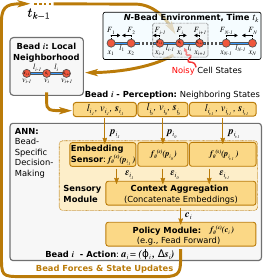}
    \caption{Schematic information-flow chart and environmental updates (chronologically following thick brown arrows) of ANN-based bead-specific decentralized decision-making implementing a system level policy that controls the locomotion of an $N$-bead microswimmer.
    The detailing ANN architecture (inspired by \cite{Hartl2024}) emphasizes an ANN's perception, $\mathbf{p}_{i_\nu}$, of bead $i$'s local neighborhood, $\nu={-1,0,1}$ (see text), followed by an embedding, $\mathbf{p}_{i_\nu}\rightarrow\varepsilon_{i_\nu}$, and concatenation layer in the sensory module that results in a bead-specific context matrix, $\mathcal{C}_i=(\varepsilon_{i_{-1}},\varepsilon_{i_0},\varepsilon_{i_{+1}})$, based on which the policy module proposes an action, $\mathbf{a}_i$, comprising the proposed force, $\phi_i$, and the cell-state update, $\Delta \mathbf{s}_i$.
    This step is performed by every bead independently at every successive time step, $t_k$, to induce an update of the state of the microswimmer at times $t_{k+1}$ by considering the regularized forces, $\phi_i\rightarrow F_i$, in the equations of motion of the $N$-bead hydrodynamic environment, $x_i(t_k)\rightarrow x_i(t_{k+1})$, and performing a noisy (c.f., red wiggly arrow) cell-state update.}
    \label{fig:methods:ann}
\end{figure}

Here, we utilize a single ANN that is independently deployed to every bead of an $N$-bead NG microswimmer to approximate a decentralized decision-making policy for autonomous locomotion of the entire virtual organism.
Thus, each ANN-augmented bead represents an agent that is immersed into a chain of $N$ single-bead agents comprising the body of an $N$-bead microswimmer (see \cref{fig:system}).
As detailed in \cref{fig:methods:ann}, the bead-specific agents of the microswimmer successively perceive the states of their respective neighboring beads and integrate this local information to initiate swimming strokes locally, following a decentralized policy that self-propels the entire microswimmer in the hydrodynamic environment.

From the perspective of \textit{Reinforcement Learning} \cite{Sutton1998}, our approach can thus be considered a trainable multi-agent system that needs to utilize local communication and decision-making to achieve a target system-level outcome \cite{Hartl2024}.
The goal is to identify a set of ANN parameters $\theta$ for the localized agents that facilitate such collective behavior, which we achieve here via evolutionary algorithms \cite{Hartl2024} (EAs), as detailed below.

Let us now specify the particular ANN architecture, and the perception (ANN input) and action (ANN output) conventions that we utilize in this contribution (as illustrated in \cref{fig:methods:ann}).

First, we define the neighborhood of a particular bead $i=1,\dotsc, N$: 
In our example of a one-dimensional, linear $N$-bead swimmer, the direct neighbors of bead $i$ are given by the beads $i\pm1$.
To address each bead in this ``$i$-neighborhood'', we introduce the index notation $i_\nu=i+\nu$ with $\nu\in\{-1, 0, 1\}$; 
$i_0$ thus addresses bead $i$ itself.

Second, we define the perception, or ANN input of bead $i$ as $\mathcal{P}_i = \{\mathbf{p}_{i_{-1}}, \mathbf{p}_{i_{0}}, \mathbf{p}_{i_{-1}}\}$, a composite matrix containing local, neighbor $i_\nu$-specific perceptions, $\mathbf{p}_{i_\nu}$, of bead $i$:
We define the neighbor-specific perception as $\mathbf{p}_{i_\nu}=(l_{i_\nu}, v_{i_\nu}, \mathbf{s}_i)$, with bead $i_\nu$-specific arm length to the neighboring beads $l_{i_\nu}=|x_i-x_{i_\nu}|\in\mathbb{R}$, bead velocity $v_{i_\nu}\in\mathbb{R}$, and an internal, vector-valued state $\mathbf{s}_i\in\mathbb{R}^{n_\mathrm{ca}}$ at time $t_k$ (see below);
out-of-bound inputs for the head and tail beads are discarded by formally setting $\mathbf{p}_{0} = \mathbf{p}_{N+1} = \mathbf{0}$ as we count $i=1,\dotsc,N$.

The internal state of a bead is inspired by the cell state of a \textit{Cellular-} \cite{vonNeumann1966}, or rather \textit{Neural Cellular Automaton} \cite{Mordvintsev2020} (NCA) that can be utilized by each bead to memorize or exchange information with neighbors.
Analogous to previous work \cite{Hartl2024}, we define the update of the internal state of a bead between two successive time steps as $\mathbf{s}_i(t_{k+1}) = (\mathbf{s}_i(t_{k}) + \Delta\mathbf{s}_i(t_k) + \mathbf{\xi}_\mathbf{s})$, where we introduced a zero-centered Gaussian noise term of STD \params{$\mathbf{\xi}_\mathbf{s}=2^{-5}$} increasing the robustness of evolved solutions \cite{Hartl2024}. 
Additionally, we clamp the elements of $\mathbf{s}_i(t_{k})$ to the interval \params{$[-1, 1]$} after each update.

Third, we here utilize a fixed ANN architecture and deploy it to every single-bead agent, as illustrated in \cref{fig:methods:ann} (see Ref.~\citenum{Hartl2024}):
we partition a bead's ANN into a sensory module, $f_\theta^{(s)}(\cdot)$, and a policy module, $f_\theta^{(c)}(\cdot)$.
The sensory module maps each neighbor-specific input separately into a respective sensor embedding, $\varepsilon_{i_\nu}(t_k)=f_\theta^{(s)}(\mathbf{p}_{i_\nu}(t_k))\in\mathbb{R}^{n_\mathrm{embd}}$, that are merged into a bead-specific context matrix $\mathcal{C}_i(t_k)=\left(\varepsilon_{i_{-1}}(t_k), \varepsilon_{i_{0}}(t_k), \varepsilon_{i_{+1}}(t_k)\right)$.
The subsequent policy module or controller ANN eventually outputs the  action of the beads, $\mathbf{a}_i(t_k)=f_\theta^{(c)}(\mathcal{C}_i(t_k)) = (\phi_i, \Delta \mathbf{s}_i)$, proposing a bead-specific force \params{$\phi_i\in[-F_0, F_0]$} (to-be regularized, $\phi\rightarrow F_i$, such that $\sum_i F_i=0$, see \cref{sec:system}) and an internal state update $\Delta \mathbf{s}_i\in\mathbb{R}^{n_\mathrm{ca}}$.

Forth, we specifically utilize a single-layer FF sensory module, $f_\theta^{(s)}(\cdot)$, with $(N_0^{(s)}=2+n_\mathrm{ca})$ input and $N_1^{(s)}=n_\mathrm{embd}$ output neurons with a $\tanh{(\cdot)}$ filter (the same network for all $3$ neighbors).
The $(3\times n_\mathrm{embd})$ context matrix, $\mathcal{C}_i$, is then flattened into a $(3\,n_\mathrm{embd})$-dimensional vector, which is processed by the policy module, $f_\theta^{(c)}(\cdot)$:
again, a single FF layer with $N_0^{(c)}=3\,n_\mathrm{embd}$ and $(N_1^{(c)}=1 + n_\mathrm{ca})$, followed by a clamping filter, $\sigma^{(c)}(\cdot)=\max(\min(\cdot, -1), 1)$.

Fifth, we use $n_\mathrm{ca}=2$ and $n_\mathrm{embd}=4$ in our simulations, resulting in $N_0^{(s)}\cdot (N_1^{(s)}+1)=20$ sensory module parameters, and $N_0^{(c)}\times (N_1^{(c)}+1)=39$ policy module parameters (accounting for the bias vectors), and thus in $N_\theta=59$ ANN parameters in total.

\FloatBarrier
\subsection{Genetic algorithm and neuroevolution of single-agent policies with collective goals}
\label{sub:methods:EA}
\textit{Genetic Algorithms} (GAs) are heuristic optimization techniques inspired by the process of natural selection.
In GAs, a set (or a population) of size $N_\mathrm{P}$, $\mathbf{X}=\{\mathbf{\theta}_1,\dotsc, \mathbf{\theta}_{N_\mathrm{P}}\}$, of sets of parameters (or individuals), $\mathbf{\theta}_i\in\mathbb{R}^{N_\theta}$, is maintained and modified over successive iterations (or generations) to optimize an arbitrary objective function (or a fitness score), ${r(\mathbf{\theta}_i):\mathbb{R}^{N_\theta}\rightarrow\mathbb{R}}$ \cite{Katoch2020, Hartl2024}.

Many \textit{Genetic}- or \textit{Evolutionary Algorithm} (EA) implementations have been proposed, which essentially follow the same biologically-inspired principles:
Starting from an initial, often random population, high-quality individuals are selected (i) from the most recent generation for reproduction, depending on their associated fitness scores. 
Based on these selected high-fitness ``parent'' individuals, new ``offspring'' individuals are sampled, e.g., by genetic recombination (ii) of two mating parents, $i,j$, by randomly shuffling the elements (or genes) of their associated parameters, schematically expressed as $\mathbf{\theta}_o=\mathbf{\theta}_i \bigoplus \mathbf{\theta}_j$.
Such an offspring's genome can be subjected to random mutations (iii), typically implemented by adding zero-centered Gaussian noise with a particular STD, $\mathbf{\xi}_\mathbf{\theta}$, to the corresponding parameters, $\mathbf{\theta}_o\rightarrow\mathbf{\theta}_o+\mathbf{\xi}_\mathbf{\theta}$.
The offspring then either replace (iv) existing individuals in the population or are discarded depending on their corresponding fitness score, $r(\mathbf{\theta}_o)$. 
In that way, the population is successively updated and is thus guided towards high-fitness regions in the parameter space, $\mathbb{R}^{N_\theta}$, over many generations of successive reproduction cycles~\cite{Ha2017blog, Hartl2024}.

Here, we utilize D. Ha's ``SimpleGA'' implementation \cite{Ha2017blog} (following steps (i)-(iv) above) to optimize the ANN parameters, $\theta$, of the single-bead agents of the here investigated $N$-bead microswimmers, see \cref{sec:system,sub:methods:ANN} and \cref{fig:system,fig:methods:ann}:
After initializing the ANN parameters of a population of size \params{$N_\mathrm{P}=128$} by sampling from a zero-centered Gaussian of STD \params{$\sigma_\mathbf{\theta}=0.1$}, we successively (i) select at each generation the best \params{$10\%$} of individuals for the reproduction cycle (ii, iii) - according to the fitness score described in \cref{sec:system} - and (iv) replace the remaining \params{$90\%$} of the population with sampled offsprings;
we fix the mutation rate in step (iii) to \params{$\mathbf{\xi}_\mathbf{\theta}=0.1$} and typically perform multiple independent GA runs for \params{$200-300$} generations (see \cref{fig:system}) each to ensure convergence of the evolved policies.
For every parameter set $\theta$ (per run, and per generation), we evaluate the fitness score as the average fitness of \params{$10$} independent episodes, each lasting for \params{$T=(400-800)$} environmental time-steps.
For every episode, we randomize the respective $N$-bead swimmer's initial bead positions as $x_i(0) \sim \mathcal{N}(\mu=i\,L_0, \sigma=R)$ drawn from a standard normal distribution centered around $\mu=i\,L_0$ with a STD of $\sigma=R$, and evaluate the episode fitness as mean center of mass velocity $v_T$ (see \cref{sec:system}).

\subsection{Swimming-gate Analysis}
\label{sub:methods:FFT}
In \cref{sec:results}, we define a $2\pi$-period governing equation $\hat l_i(t) = f_\ell\left((t - i\,\bar\tau)\,\bar\omega + \phi\right)$ for the actual arm lengths $l_i(t_k)$ for both \typeAB{} $N$-bead microswimmers as a function of the mean angular velocity $\bar\omega=\bar\omega(N)$ and the mean neighbor arm cross-correlation time $\bar\tau=\bar\tau(N)$.
For all evolved \typeAB{} microswimmer policies utilized in \cref{fig:results:generalize}~(A,B), we thus evaluate the corresponding mean angular velocity as $\bar\omega=\frac{1}{N-1}\sum_{i=1}^{N-1}\omega_i$ by averaging the most dominant angular velocities $\omega_i$ extracted for every arm length $l_i(t)$ of a particular swimmer realization via Fourier transformation.
We further define $\bar\tau=\frac{1}{N-1}\sum_{i=1}^{N-1}\tau_i$, 
with $\tau_i$ being the optimal time delay between neighboring arm lengths $l_i(t)$ and $l_{i+1}(t+\tau_i)$ maximizing the overlap
${\phantom{|}}\frac{d}{d\tau}\int_0^T l_i(t)l_{i+1}(t+\tau)\,dt|_{\tau=\tau_i} = 0$.


\section*{Acknowledgements}
We thank Sebastian Risi and Santosh Manicka for helpful discussions. 
BH gratefully acknowledges an APART-MINT fellowship from the Austrian Academy of Sciences. 
ML gratefully acknowledges support via Grant 62212 from the John Templeton Foundation.
The computational results presented have been achieved (in part) using the Vienna Scientific Cluster 5.

\section*{Author contributions}
BH and AZ designed the study and wrote the paper.
BH developed the code and performed simulations.
BH and AZ analyzed results.
ML discussed results, and reviewed and commented on the manuscript.

\section*{Author declarations section}
The authors have no conflicts to disclose.

\section*{Data availability statement}
The data that support the plots within this paper and other ﬁndings of this study are available from the corresponding author upon request.


\section*{References}
\bibliography{references,RL-swim}

\end{document}